\documentstyle[psfig,12pt,aaspp4]{article}
\def\kms{km~s$^{-1}$}

\def\be{\begin{equation}}
\def\ee{\end{equation}}
\def\about{$\sim$}
\def\etal{{\it et al.}}
\def\h{$h^{-1}$}

\begin{document}
\hskip 3.5in{\hskip 10pt \date{1 August 1998}}
\title{SEEKING THE LOCAL CONVERGENCE DEPTH. THE ABELL CLUSTER DIPOLE FLOW TO 200\h\ MPC.}
 
\author {DANIEL A. DALE,\altaffilmark{1} RICCARDO GIOVANELLI, AND MARTHA P. HAYNES,}\affil{Center for Radiophysics and Space Research and National Astronomy and Ionosphere Center, Cornell University, Ithaca, NY 14853}

\author {LUIS E. CAMPUSANO}
\affil{Observatorio Astron\'{o}mico Cerro Cal\'{a}n, Departamento de 
Astronom\'{\i}a, Universidad de Chile, Casilla 36-D, Santiago, Chile}

\author {EDUARDO HARDY}
\affil{National Radio Astronomy Observatory, Casilla 36-D, Santiago, Chile}

\author {STEFANO BORGANI}
\affil{INFN, Sezione di Perugia, c/o Dipartimento di Fisica dell'Universit\`{a}, via A. Pascoli, I-06100 Perugia, Italy}

\altaffiltext{1}{Now at IPAC, California Institute of Technology 100-22, Pasadena, CA 91125}

\begin{abstract}
We have obtained new Tully-Fisher (TF) peculiar velocity measurements for 52 Abell galaxy clusters distributed throughout the sky between \about 50 and 200\h\ Mpc.  The measurements are based on $I$ band photometry and optical rotation curves for a sample of 522 spiral galaxies, from which an accurate TF template relation has been constructed.  Individual cluster TF relations are referred to the template to compute cluster peculiar motions.

The reflex motion of the Local Group of galaxies is measured with respect to the reference frame defined by our cluster sample and the distant portion of the Giovanelli \etal\ (1998a) cluster set.  We find the Local Group motion in this frame to be 565$\pm113$ \kms\ in the direction $(l,b)=(267^\circ,26^\circ) \pm10^\circ$ when peculiar velocities are weighted according to their errors.  After optimizing the dipole calculation to sample equal volumes equally, the vector is $509\pm195$ \kms\ towards $(255^\circ,33^\circ)\pm22^\circ$.  Both solutions agree, to within 1$\sigma$ or better, with the Local Group motion as inferred from the cosmic microwave background (CMB) dipole.  Thus, the cluster sample as a whole moves slowly in the CMB reference frame, its bulk flow being at most 200 \kms.
\end{abstract}


\section {INTRODUCTION}
The motion of the Local Group of galaxies is imprecisely known.  Using the COBE result for the motion of the Sun in the CMB reference frame (Kogut \etal\ 1993) and the recipe of de Vaucouleurs \etal\ (1976) to convert heliocentric velocities to the Local Group frame yields a Local Group velocity of ${\bf V}_{\rm lg-cmb}=620\pm22$ \kms\ towards $(l,b)=(271^\circ,29^\circ)\pm3^\circ$.  The volume over which the coherent motion represented by ${\bf V}_{\rm lg-cmb}$
persists is unknown.  It can be estimated, however, by measuring the reflex motion, with respect to the Local Group, of spherical shells of increasing radii (Giovanelli \etal\ 1998b; G98b).  The ``convergence depth'' is the distance at which the shells' dipole motions begin to mirror the Local Group motion. Alternatively, the distance at which the shells' {\it bulk motions}, the dipole motions in the CMB reference frame, approach rest defines convergence.  Using Tully-Fisher (TF) distances for a large sample of field galaxies out to 95\h\ Mpc, G98b find the convergence depth to be of order 60\h\ Mpc.  This result is strengthened by the work of Giovanelli \etal\ (1998a; G98a), who find evidence for a small bulk flow using TF distances on a sample of clusters between 30 and 92\h\ Mpc.  These efforts, however, do not preclude non-zero bulk motions on scales as large as those claimed by Lauer \& Postman (1994; LP).  In this letter we tackle the issue of large--scale bulk motions by measuring the reflex motion, with respect to the Local Group, of a deep, all--sky sample of 52 Abell clusters that extends to \about 200\h\ Mpc.  To fortify our results, we merge to that sample the distant half of the G98a sample.  In the following section we highlight the general properties of the data.  Section \ref{sec:dipole} deals with dipole calculations of the cluster peculiar velocity sample and Section \ref{sec:ZoA} covers the effect the Zone of Avoidance has on such computations.  We review and discuss our results in Section \ref{sec:summary}.

\section{THE ABELL CLUSTER INERTIAL REFERENCE FRAME}
The TF data for this project have been collected over the last four years.  Installments of the data have been presented previously (Dale \etal\ 1997, 1998), with the third and final installment forthcoming.  A description of the entire sample's properties, the construction of the $I$ band TF template, the sample biases and their corrections, and the individual cluster peculiar velocities will appear in a separate work.  We briefly summarize the data here. 

\begin{figure}[!ht]
\centerline{\psfig{figure=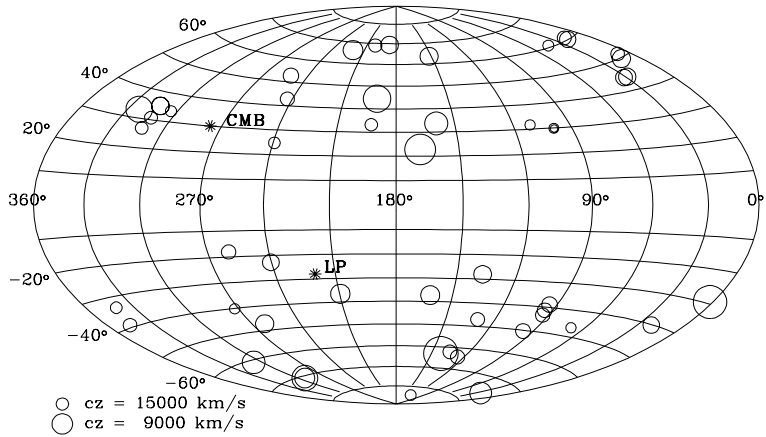,width=4in,bbllx=30pt,bblly=234pt,bburx=585pt,bbury=560pt,clip=t}}
\caption[The SCII Sample]
{\ The all-sky distribution of the SCII sample in Galactic coordinates.  The  circles are drawn inversely proportional to the cluster redshifts.  The two examples in the lower left give the scale.  Asterisks mark the apices of the motion of the Local Group with respect to the CMB and the LP cluster inertial frame.}
\label{fig:aitoff_sample}
\end{figure}

The sample consists of an all--sky collection of 52 rich Abell clusters between \about 50 and 200\h\ Mpc (hereafter the `SCII' sample).  Figure \ref{fig:aitoff_sample} displays the cluster distribution in Galactic coordinates.  
%%%%%%%%%%%%%%%%%%%%%%%%%%%%%%%%%%%%%%
%%%%%%%%%%%%%%%%%%%%%%%%%%%%%%%%%%%
The symbol sizes are scaled inversely proportional to the cluster redshifts.  The average of the cluster CMB velocities is 12,100 \kms.  TF peculiar motions have been estimated for these clusters based on $I$ band photometry and optical rotation curve work for 522 spiral galaxies.  The number of galaxies with TF data and their dispersion determine the accuracy of a given cluster's peculiar motion estimate.  For our sample, the average cluster peculiar velocity uncertainty is \about 6\% of the cluster distance.  

The sky and redshift distributions of our cluster sample play important roles in this study.  Adequate sky coverage is necessary for any dipole measurement and the redshift distribution must be tailored to test claims of bulk flows on large scales.  Our sample has been selected according to these precepts, but it is of course useful to consider supplementing our data with peculiar velocity measurements of other clusters.  The distant members (c$z>4500$ \kms) of the G98a `SCI' sample afford such an opportunity.  Our data represent an extension of the SCI data to higher redshifts so we have taken care to process the data in a manner consistent with the data reductions for the SCI program.  To ensure uniform sky coverage and a redshift overlap with the SCII sample, we restrict the following dipole computations to involve only the distant half of the SCI sample, the 12 clusters beyond 45\h\ Mpc.  We shall refer to the combined sample of 64 SCI+SCII clusters as the SC sample.  The average redshift of the SC sample in the CMB frame is anywhere from 7500 to 11,500 \kms, depending on the weighting scheme.  We point out that this sample depth matches that of the LP data set, a sample for which the local bulk flow was found to be on the order of 700 \kms.

\section{DIPOLES AND BULK FLOWS}
\label{sec:dipole}

\subsection{The Abell Cluster Dipole}
\begin{figure}[!ht]
\centerline{\psfig{figure=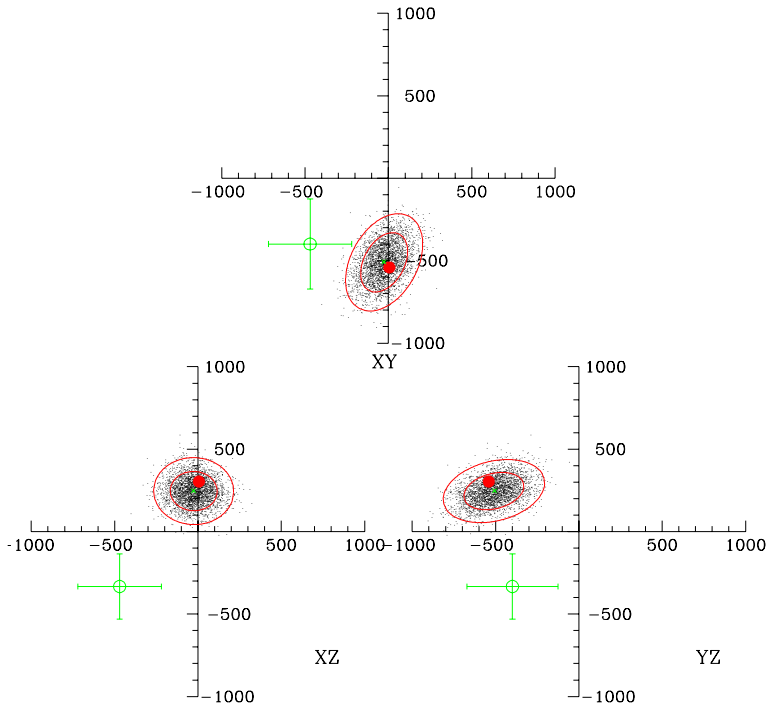,width=4in,bbllx=15pt,bblly=152pt,bburx=579pt,bbury=676pt,clip=t}}
\caption[Dipole Confidence Ellipses I]
{\ Galactic Cartesian coordinate representation of the motion of the Local Group with respect to the Abell cluster inertial reference frame.  The small dots plotted are Monte Carlo realizations of the peculiar velocity data set, and the ellipses drawn represent the 1$\sigma$ and 2$\sigma$ confidence regions of the dipole.  The large filled circles indicate the motion of the Local Group with respect to the CMB and the open circles show the dipole motion with respect to the LP reference frame, along with the quoted uncertainty.}
\label{fig:ellipse}
\end{figure}

The reflex motion of the Local Group with respect to the SC sample, ${\bf V}_{\rm lg-sc}$, follows from a chi-squared minimization of the dipole vector with respect to the cluster sample's peculiar velocity field {\it in the Local Group frame}, i.e. the ${\bf V}_{\rm lg-sc}$ that minimizes
\be
\chi^2 = \sum_{i=1}^N {1 \over s_i}
\Big({V_i + {\bf \hat r}_i \cdot {\bf V}_{\rm lg-sc} \over \epsilon_i}\Big)^2
\label{eq:dipole}
\ee
where $\hat {\bf r}_i$ is the unit vector in the direction of the $i^{\rm th}$ 
cluster, and $V_i$ and $\epsilon_i$ are the cluster's peculiar velocity and its error.  The factor $1/s_i$ allows us to vary the weighting scheme; we assume $1/s_i$=1 in what immediately follows, and we will henceforth call this ``Case $a$.''  A straightforward minimization of this equation yields ${\bf V}_{\rm lg-sc}=(V_x,V_y,V_z)=(-29,-507,247)$ \kms\ in Galactic Cartesian coordinates, or alternatively, 565 \kms\ towards $(l,b)=(267^\circ,26^\circ)$.  A robust approach to estimating the uncertainty in the dipole is to utilize the characteristics of the cloud of solutions generated from Monte Carlo simulations.  In these simulations we generate a Monte Carlo trial sample by adding random Gaussian deviates of the peculiar velocity errors to the original peculiar velocities.  The dipole solution is calculated, and the process is repeated for a large number ($10^4)$ of realizations.  The dispersions in the dipole components, (93,119,80) \kms, imply an error-bias corrected dipole amplitude of $538\pm113$ \kms\ (error bias systematically inflates the dipole amplitude: $V^2_{\rm cor} = V_x^2+V_y^2+V_z^2- \epsilon_x^2-\epsilon_y^2-\epsilon_z^2$).  Figure \ref{fig:ellipse} displays the error cloud projected onto the three planes of the Galactic Cartesian coordinate system.  
%%%%%%%%%%%%%%%%%%%%%%%%%%%%%%%%%%%%%%
%%%%%%%%%%%%%%%%%%%%%%%%%%%%%%%%%%%%%%
We can use this error cloud to better understand the statistical likelihood domains of the measurement; the contours plotted enclose 68\% and 95\% of the points, respectively representing 1$\sigma$ and 2$\sigma$ confidence regions.  The contours derive from two-dimensional Gaussian fits to the Monte Carlo distribution.  The filled circles indicate the dipole motion of the Local Group with respect to the CMB, whereas the open circles with the accompanying error bars show the value as determined by LP.  

A $\Delta \chi^2$ analysis can be another approach to understanding the dipole result.  The difference in $\chi^2$ using our solution and one with a null reflex Local Group motion is $\Delta \chi^2=51.4$ (or $\Delta \chi^2=38.5$ if we account for the errors on the CMB dipole).  For a three parameter fit this implies we can guarantee non-zero Local Group motion with 99.999\% confidence.  More importantly, we can test for agreement with the COBE solution for Local Group motion, ${\bf V}_{\rm lg-cmb} = (6,-541,303)$ \kms.  In this case $\Delta \chi^2=1.4$; our results for the Local Group motion agree with that of COBE to well within 1$\sigma$ --- the COBE solution lies within our 68\% confidence region.  Conversely, $\Delta \chi^2=17.8$ in a comparison between our solution and that of LP (computed including the LP error estimate); we disagree with the LP result to better than 99.9\% (a 3.5$\sigma$ difference).

Table 1 summarizes the dipole results, parenthetically including the dipole amplitudes corrected for error biasing.  An indication of the effective sample depth is included, computed by averaging cluster recessional velocities weighted by $1/(s_i \epsilon_i^2)$.  The number of clusters considered in each computation, $N_{\rm cl}$, exceeds the number of SC clusters (64) when artificial clusters from the Zone of Avoidance are included, as described later.  Finally, we have also computed the dipole after correcting for a homogeneous Malmquist bias, in the form advocated by G98b.  This correction alters results little; it is calculated to estimate the amplitude and potential impact of the effect.  It is likely that a more complicated, but harder to quantify, inhomogeneous bias correction is necessary.  However, the variance between the inhomogeneous and the homogeneous forms of the correction is likely to be a second order effect; either form impacts little the final result.

%%%%%%%%%%%%%%%%%%%%%%%%%%%%%%
\begin{table}[!ht]
\centerline{Table 1.  Cluster Dipole}
%\caption[Cluster Dipole]{Cluster Dipole}
\def\Nc{$N_{\rm cl}$}
\def\sia{Case $a$: $s_i \equiv 1$ \hskip 2cm $\left< {\rm c}z \right> \sim 7400$ \kms}
\def\sib{Case $b$: $s_i\epsilon_i^2 \propto 1/\sqrt{N_{\rm i}}$ \hskip 0.7cm $\left< {\rm c}z \right> \sim 10,600$ \kms}
\def\p{$\pm$}
\def\note{c$z_{\rm tf}>$4500 \kms}
\def\descrb{incl. HMBC\tablenotemark{a}}
\def\descrc{incl. ZoA clusters\tablenotemark{b}}
\def\descrd{incl. ZoA clusters\tablenotemark{c}}
\def\descre{excl. A3667}

\begin{center}
\begin{tabular}{lccccc} \tableline \tableline
SCI+SCII & \Nc &    V           & ($l,b$)      \\
\note    &     &   \kms         & $^\circ$     \\
\tableline
\sia     &     &               &               \\
         & 64  & 565(538)\p113 & (267,+26)\p10 \\
\descrb  & 64  & 564(538)\p113 & (265,+27)\p11 \\
\descrc  & 80  & 447(417)\p101 & (273,+39)\p15 \\
\descrd  & 80  & 550(526)\p104 & (259,+28)\p11 \\
\descre  & 63  & 564(537)\p114 & (266,+26)\p10 \\
         &     &               &               \\
\sib     &     &               &               \\
         & 64  & 496(393)\p205 & (269,+28)\p22 \\
\descrb  & 64  & 593(510)\p208 & (264,+25)\p17 \\
\descrc  & 80  & 354(254)\p145 & (272,+43)\p32 \\
\descrd  & 80  & 492(427)\p152 & (249,+28)\p19 \\
\descre  & 63  & 509(411)\p195 & (255,+33)\p22 \\
\tableline
\end{tabular}
\end{center}
\tablenotetext{a}{homogeneous Malmquist bias correction}
\tablenotetext{b}{Artificial ZoA clusters included, with null average peculiar velocity in Local Group frame}
\tablenotetext{c}{Artificial ZoA clusters included, with null average peculiar velocity in LP frame}
\label{tab:dipole}
\end{table}
%%%%%%%%%%%%%%%%%%%%%%%%%%%%%%

\subsection{The Dipole Under Equal Volume Weighting}

\begin{figure}[!ht]
\centerline{\psfig{figure=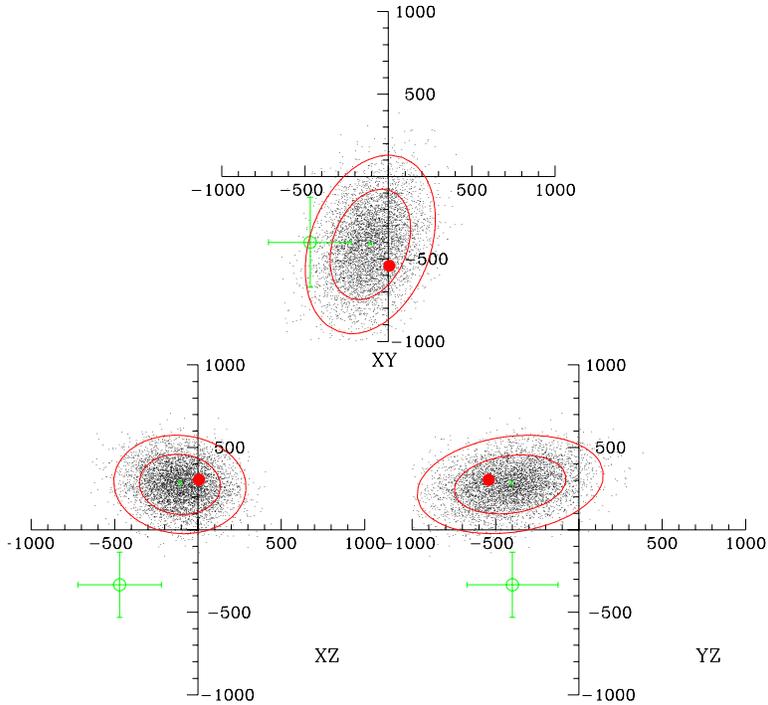,width=4in,bbllx=15pt,bblly=152pt,bburx=579pt,bbury=676pt,clip=t}}
\caption[Dipole Confidence Ellipses II]
{\ Similar to Figure \ref{fig:ellipse}, but for equal volume weighting: $1/(s_i \epsilon_i^2) = 1$.}
\label{fig:ellipse2}
\end{figure}

The previous formalism does not weigh equal volumes equally, but rather tends to weigh nearby regions more heavily.  This unequal volume weighting exists because the completeness of our cluster sample and the accuracy of our peculiar velocities fall with increasing distance.  An exponential fit to the completeness function of the SCI+SCII sample follows $\exp(-{\rm c}z [{\rm km/s}]/4700)$.  To accomplish equal volume weighting, or more precisely {\it equal spherical shell weighting}, we construct the factor $1/s_i$ in Equation \ref{eq:dipole} to account for the characteristics of our cluster selection function.  The resulting combination of our cluster selection function and the peculiar velocity error distribution yields a roughly flat weighting with redshift: $1/(s_i \epsilon_i^2) \propto 1/\sqrt{N_i}$, hereafter referred to as ``Case $b$,'' where $N_i$ is the number of galaxies used in the $i^{\rm th}$ cluster.  Results for the dipole computations using this fair weighting scheme are also listed in Table 1.  We again find agreement with the COBE solution, though the associated errors are admittedly larger -- giving similar weights to near and far clusters inevitably increases the dipole uncertainty since distant clusters have less accurate peculiar velocities.  Results are also listed for the dipole computation when the cluster Abell 3667 is excluded, an object with a highly uncertain peculiar velocity estimate, as it is based on TF measurements of only four galaxies.  By performing jackknife and statistical tests of the Case $b$ dipole, we find this cluster to unduly alter the results (its presence in Case $a$ solutions makes little difference due to its large peculiar velocity error).  We thus conservatively exclude this cluster from further Case $b$ results.  The Monte Carlo realizations for the Case $b$ dipole (Figure \ref{fig:ellipse2}) are in good agreement with the CMB dipole.
%%%%%%%%%%%%%%%%%%%%%%%%%%%%%%%%%%%%%%
%%%%%%%%%%%%%%%%%%%%%%%%%%%%%%%%%%%%%%

\subsection{The Bulk Flow}
We use these results in concert with the estimate of the Local Group motion in the CMB frame to ascertain the motion of the cluster inertial frame with respect to the CMB,
\be
{\bf V}_{\rm sc-cmb} \equiv {\bf V}_{\rm lg-cmb} - {\bf V}_{\rm lg-sc}.
\ee
We find the bulk flow amplitude for the two cases $1/s_i=1$ and $1/(s_i \epsilon_i^2) \propto 1/\sqrt{N_i}$ to be $75\pm92$ \kms\ and $178\pm196$ \kms; corrected for error biasing, the amplitudes reduce to a null bulk flow.  The inertial frame defined by the distant SCI+SCII clusters is determined to be at rest with respect to the CMB fiducial rest frame.  We note that the above estimates for the local Universe's bulk flow amplitude match well what is expected from theoretical considerations (e.g. Feldman and Watkins 1994; Strauss \etal\ 1995; Moscardini \etal\ 1996).

\section{THE EFFECT OF THE ZONE OF AVOIDANCE}
\label{sec:ZoA}
The region of the sky that is obscured by the plane of the Galaxy is rather large even in the $I$ band; the Zone of Avoidance (ZoA) extends to about $\pm20^{\circ}$ in Galactic latitude, or alternatively, it covers about a third of the sky.  We have abstained from working at low Galactic latitudes because Galactic extinction corrections there are highly uncertain.  We first turn to Monte Carlo simulations to gauge the effect of the ZoA on bulk motion measures.  For each of $10^4$ trial realizations of the dipole we generate 15--20 artificial clusters between $-20^\circ<b<20^\circ$ that mimic the overall characteristics of the observed cluster sample.  Table 1 includes the average results of these simulations, with the quoted errors deriving from the $1\sigma$ dispersions.  The amplitudes and apices agree less well with ${\bf V}_{\rm lg-cmb}$, but we emphasize that the artificial clusters added have, on average, zero peculiar velocity in the Local Group frame in one case or they have been assigned, on average, the LP dipole motion in another case.  This exercise is carried out to gauge, very conservatively, the maximum potential effect of the unsampled ZoA region.

%\subsection{Probing ZoA Effects With Numerical Simulations}

Numerical simulations based on the truncated Zel'dovich approximation kindly provided by S. Borgani allow us to further test the impact of the ZoA and the effects of cosmic scatter on our dipole calculations.  Such simulations allow one to gauge uncertainties introduced by blanking large swaths of the sky.  In other words, comparisons can be made between dipoles that sample the entire sky and those that are limited by a ZoA.  Moreover, simulated peculiar velocities have no ``observational errors.''  Thus, we can test for different effects to a higher accuracy than would otherwise be attainable.  As described in Moscardini \etal\ (1996), the simulations use the Zel'dovich approximation to mimic gravitational dynamics.  The simulations are carried out on a $256^3$ grid, one particle per grid point, within a cube of 960\h\ Mpc sides, a region large enough to contain all the waves responsible for motions on scales of 200\h\ Mpc.  A variety of cosmological models are simulated.

To enable a fair comparison with our cluster observations, we cull the number of density peaks to a density similar to that found in the Abell/ACO catalog: $n_{\rm cl} \simeq 2 \times 10^{-5} (h^{-1} {\rm Mpc})^{-3},$ or an average cluster-cluster spacing of $\simeq 40$\h\ Mpc.  We also only calculate a cluster dipole for ``Local Group-like'' observers.  In other words, $|V_{\rm lg}|=620\pm44$ \kms\ for a top-hat sphere of radius 7.5\h\ Mpc, and a matter overdensity range of $-0.2 \leq \delta \rho /\rho \leq 1$.  Such criteria are satisfied by $10^2$ to $10^3$ grid points, depending on the cosmological model.

The dipole calculation is executed three times for each Local Group-like observer using a random sample of clusters whose space distribution is fashioned after the combined SCI+SCII cluster sample.  The first computation tests only for the effect of cosmic variance in the peculiar velocity field and thus does not involve any sort of sky masking.  The apex of this first dipole computation can also serve as a working reference point for the ``true'' dipole.  The second calculation includes a randomly placed ZoA for each observer, whereas the third calculation involves a ZoA mask that is oriented, with respect to the true apex, in the same manner we observe the plane of the Milky Way with respect to the apex of our own Local Group motion in the CMB frame.  The variance of the dipole solution for the set of Local Group-like observers gives us an idea of how well we can recover dipole flows.  The majority of the simulations indicate that the measurement of 60--70 cluster cosmic motions limit bulk flow measurements to an accuracy no better than \about 100--150 \kms; the presence of a ZoA further confuses dipole measurements by \about 60 \kms, resulting in an overall uncertainty of \about 120--160 \kms, consistent with the errors in Table 1.

\section{CONCLUSIONS}
\label{sec:summary}
Our goal is to conclusively test claims of high amplitude, large scale bulk motions using Tully-Fisher redshift-independent distances to galaxy clusters.  In collaboration with the accurate SCI peculiar velocities, we use our sample of 52 cluster peculiar velocities to accomplish our goal, and in the process we establish a reliable upper limit to the local convergence depth.  We briefly recap here our findings.

$\bullet$ The reflex motion of the Local Group is calculated with respect to the Abell cluster inertial reference frame represented by the distant SCI+SCII sample.  Upon weighting peculiar velocities by their error estimates, we determine the Local Group motion to be 565$\pm113$ \kms\ in the direction $(l,b)=(267^\circ,26^\circ)\pm10^\circ$.  Employing a different weighting scheme, one where equal volumes are treated equally, yields $509\pm195$ \kms\ towards $(255^\circ,33^\circ)\pm22^\circ$.  Both results agree with the CMB dipole.  The bulk motion of the SC sample in the CMB reference frame is at most 200 \kms.

$\bullet$ We employ a variety of numerical simulations to test the statistical validity of our results.  We find that bulk flow measures not derived from a truly all-sky sample, but rather those from samples suffering from a dearth of objects at low galactic latitudes such as for our survey, can be reliable.  This fact holds for a variety of cosmological scenarios.

$\bullet$ We have applied a suite of statistical tests to our dipole 
calculations.  Bootstrap and subset resamplings in addition to jackknife tests indicate our results are robust.

$\bullet$ We find evidence supporting the small convergence depth of order 60\h\ Mpc found by Giovanelli and coworkers.  They use two independent samples of TF distances, one comprised of 24 clusters out to 92\h\ Mpc (G98a) and the other made up of a large sample of spiral field galaxies covering a similar distance regime (the `SFI' sample; G98b).  With respect to spherical shells of increasing radii, they find the reflex motion of the Local Group increasingly mirrors the CMB dipole.  Our result shows that the bulk flow of the volume subtended by the SC sample, of effective depth \about 100\h\ Mpc, is less than 200 \kms, thus confirming convergence is maintained beyond the limits of the SCI and SFI samples.

\acknowledgements
We would like to thank the referee for providing useful comments.  The results presented here are based on observations carried out at the Palomar Observatory (PO), at the Kitt Peak National Observatory (KPNO), at the Cerro Tololo Inter--American  Observatory (CTIO), and the Arecibo Observatory, which is part of the National Astronomy and Ionosphere Center (NAIC).  KPNO and CTIO are operated by Associated Universities for Research in Astronomy and NAIC is operated by Cornell University, all under cooperative agreements with the National Science Foundation.  The Hale telescope at the PO is operated by the California Institute of Technology under a cooperative agreement with Cornell University and the Jet Propulsion Laboratory.  This research was supported by NSF grants AST94-20505 and AST96--17069 to RG and AST95-28960 to MH.  LEC was partially supported by FONDECYT grant \#1970735.

%\newpage

%\newpage

\begin{thebibliography}{dum}

\bibitem[]{} 
Dale, D.A., Giovanelli, R., Haynes, M.P., Scodeggio, M., Hardy, E., \& Campusano, L. 1997, \aj, 114, 455

\bibitem[]{} 
Dale, D.A., Giovanelli, R., Haynes, M.P., Scodeggio, M., Hardy, E., \& Campusano, L. 1998, \aj, 115, 418

\bibitem[]{}
de Vaucouleurs, G., de Vaucouleurs, A., \& Corwin, H.G. 1976, {\it Second Reference Catalogue of Bright Galaxies}. New York: Springer

\bibitem[]{}
Feldman, H.A. \& Watkins, R. 1994, \apjl, 430, L17

\bibitem[]{}
Giovanelli, R., Haynes, M.P., Salzer, J J., Wegner, G., Da Costa, L.N., \& 
Freudling, W. 1998, \aj\ (to appear) (G98a)
 
\bibitem[]{}
Giovanelli, R., Haynes, M.P., Freudling, W., Da Costa, L.N., Salzer, J J., \& Wegner, G. 1998, \apjl, 505, L91 (G98b)

\bibitem[]{}
Kogut, A. \etal\ 1993, \apj, 419, 1

\bibitem[]{}
Lauer, T. \& Postman, M. 1994, \apj, 425, 418 (LP)

\bibitem[]{} 
Lynden-Bell, D. \& Lahav, O. 1988, in {\it Large Scale Motions in the Universe: A Vatican Study Week}, ed. by V.C. Rubin and G.V. Coyne, S.J., Princeton, U. Press, Princeton, p. 199

\bibitem[]{}
Moscardini, L., Branchini, E., Tini--Brunozzi, P., Borgani, Plionis, M., \& Coles, P. 1996, \mnras, 282, 384

\bibitem[]{}
Strauss, M.A., Cen, R., Ostriker, J.P., Lauer, T.R., \& Postman, M. 1995, \apj, 444, 507

\bibitem[]{}
Yahil, A., Tammann, G.A., \& Sandage, A. 1977, \apj, 217, 903

\end{thebibliography}
\end{document}